\documentclass[%
reprint, showpacs,
 onecolumn, notitlepage,
 amsmath,amssymb,
 aps,
 prl, superscriptaddress,
]{revtex4-1}

\usepackage{graphicx}
\usepackage{dcolumn}
\usepackage{bm}
\usepackage{enumerate}

\usepackage{color}
\usepackage{hyperref}

\usepackage{booktabs} 
\usepackage{array} 
\usepackage{paralist} 
\usepackage{verbatim} 

\usepackage{fancyhdr} 
\usepackage{amsmath}
\usepackage{graphicx}
\usepackage{amsthm}

\theoremstyle{definition}

\DeclareMathOperator{\Tr}{Tr}

\newcommand{\be}{\begin{equation}}
\newcommand{\ee}{\end{equation}}
\newcommand{\bea}{\begin{eqnarray}}
\newcommand{\eea}{\end{eqnarray}}

\newcommand{\dS}{\mathrm{dS}}

\begin{document}

\preprint{CALT 2017-006}

\title{Quantum Circuit Cosmology:
\\ The Expansion of the Universe Since the First Qubit
}

\author{Ning Bao}
\email{ningbao75@gmail.com}
\affiliation{Walter Burke Institute for Theoretical Physics, California Institute of Technology, Pasadena, CA 91125, USA}
\affiliation{Institute of Quantum Information and Matter, California Institute of Technology, Pasadena, CA 91125, USA}
\author{ChunJun Cao}
\email{ccj991@gmail.com}
\affiliation{Walter Burke Institute for Theoretical Physics, California Institute of Technology, Pasadena, CA 91125, USA}
\author{Sean M.\ Carroll}
\email{seancarroll@gmail.com}
\affiliation{Walter Burke Institute for Theoretical Physics, California Institute of Technology, Pasadena, CA 91125, USA}
\author{Liam McAllister}
\email{liam.mcallister@gmail.com}
\affiliation{Department of Physics, Cornell University, Ithaca, NY 14853, USA}

\begin{abstract}
We consider cosmological evolution from the perspective of quantum information.
We present a quantum circuit model for the expansion of a comoving region of space,
in which initially-unentangled ancilla qubits become entangled as expansion proceeds.
We apply this model to the comoving region that now coincides with our Hubble volume, taking
the number of entangled degrees of freedom in this region to be proportional to the de Sitter entropy.
The quantum circuit model is applicable for at most 140 $e$-folds of inflationary and post-inflationary expansion: we argue that no geometric description was possible before the time $t_1$ when our comoving region was one Planck length across, and contained one pair of entangled degrees of freedom.
This approach could provide a framework for modeling the initial state of inflationary perturbations.

\end{abstract}

\maketitle

\section{Introduction}

Predictions of inflationary cosmology \cite{Guth:1980zm,Linde:1981mu,Albrecht:1982wi} are generally derived in the framework of quantum fields evolving in a classical background spacetime.  While this approach has had empirical success, it raises an important conceptual problem: degrees of freedom are represented as modes of fixed comoving wavelength, and
as space expands, modes with wavelengths less than the Planck length are stretched to be super-Planckian, and so become visible in a long-wavelength effective description.  One manifestation of this issue is the trans-Planckian problem (see e.g.~\cite{Martin-Brandenberger:2001,Hogan:2002xs,Easther:2002xe,Kaloper:2002uj,Mathur:2003ez,Collins:2005nu,Brandenberger-Martin:2013}), which asks whether newly-appearing modes are in a state other than the usual Bunch-Davies vacuum \cite{Bunch-Davies:1978}, and if so, how this affects the predictions of inflation.

Our concern in this paper is with a deeper problem: not the quantum state of modes that are initially trans-Planckian, but the very nature and existence of such modes.
In the context of
quantum field theory in curved spacetime, where the dimensionality of Hilbert space is infinite, it is possible in principle to imagine a limitless store of zero-energy modes initially frozen into their vacuum states, which become dynamical when their wavelengths grow longer than the Planck length.
But is this infinite supply of degrees of freedom physically meaningful?
In this note we confront this problem from the perspective of the emergence of spacetime
from quantum entanglement \cite{Ryu:2006bv,vanRaamsdonk:2010,Hartman:2013qma,Faulkner-et.al.:2013,Jacobson:1995ab,Jacobson:2015,Banks:2011av,Banks:2015iya,Cao-Carroll-Michalakis:2016}.

We suggest that each finite-sized comoving region of space is described by a finite number of quantum degrees of freedom, so the supply of new modes is not limitless.  Concretely, we posit that a finite comoving region of space can be described by a density matrix associated with a Hilbert space ${\cal{H}}$ of fixed, finite dimension $D$.  A convenient, though logically inessential, representation takes ${\cal{H}}$ to be the tensor product of $n$ qubit degrees of freedom, so that $D = 2^n$.
These degrees of freedom include both those describing space itself, and the modes of an emergent field theory on wavelengths much larger than the Planck scale.

\begin{figure}
  \includegraphics[width=0.50\textwidth]{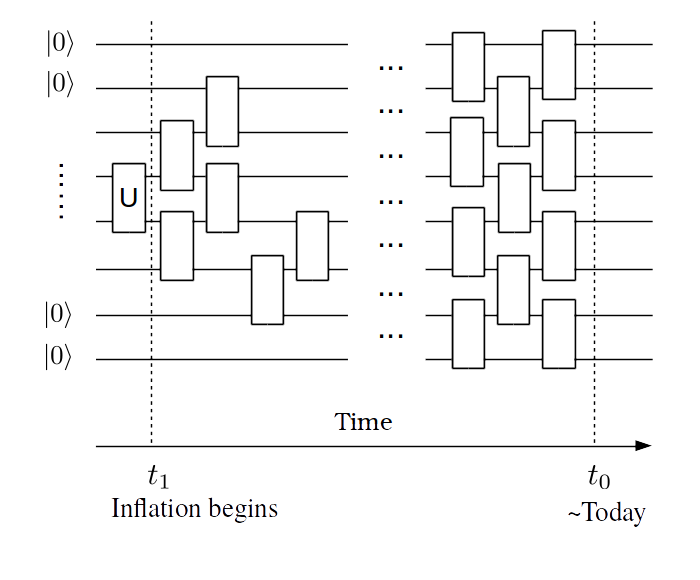}
  \caption{A schematic for a cosmological quantum circuit. The ancillary qubits are initialized in $|0\rangle$ states, and the boxes are unspecified unitary gates.}
  \label{fig:circuit}
\end{figure}

As a toy model for the evolution of a fixed comoving region ${\cal C}$, we propose a simple quantum circuit.
(Our approach thus bears a family resemblance to the proposal that the universe can be thought of as a quantum computer \cite{Lloyd:2005js,Markopoulou:2012np}.)
A quantum circuit consists of a network of quantum gates, each of which performs a unitary transformation on the basic factors of the Hilbert space of a quantum system, which we have taken to be qubits.  This yields a convenient representation of the evolution of the system.
At any time $t$, we can divide the $n$ degrees of freedom in ${\cal{H}}$ into a number $n_e(t)$ that are entangled with each other (and whose entanglements are responsible for the spacetime structure), and a number $n_u(t)$ that are not entangled with anything:
\begin{equation}
n = n_e(t) + n_u(t)\,.
\end{equation}
The unentangled degrees of freedom can be thought of as ``ancilla'' qubits.
These  are initially not entangled with each other, nor with the degrees of freedom describing other regions.
In our model, as space expands and the physical size of ${\cal C}$ increases, no new degrees of freedom are brought into existence.
Instead, more and more of the ancilla qubits become entangled with those that are already part of the spacetime structure.
The fundamental gate in our quantum circuit entangles an ancilla qubit with the rest of the circuit; this is interpreted as a small amount of expansion.
See Fig.~\ref{fig:circuit}.

We will apply this picture to the history of our comoving patch, i.e.~the comoving volume that now coincides with our Hubble volume.
For a Hubble volume in de Sitter space we argue that $D \approx e^S$, with $S$ the de Sitter entropy, so that $n  \approx n_e \approx S/\ln 2$.
Our own comoving patch therefore contains $n \approx 10^{122}$ entangled degrees of freedom today.  In our past, when our comoving region was smaller, many of these degrees of freedom were not yet entangled.

This picture provides a candidate description of the quantum state of our comoving region at very early times.
If inflation lasted for just the minimal number of $e$-folds necessary to solve the horizon problem, then at the start of inflation our comoving region was approximately a Hubble volume.  However, if inflation lasted slightly longer than this (as quantified below), then sufficiently early in inflation the diameter of our comoving region was the Planck length $\ell_p$.
A semiclassical description of quantum fields in this region is problematic, because the wavelengths of such modes are $< \ell_p$.
In most approaches to the trans-Planckian problem, the underlying spacetime is taken to be smooth, and ambiguities associated to modes on this background are  addressed by imposing a cutoff prescription.

In our picture, in contrast, even the notion of a classical metric ceases to make sense for a region of Planckian size: we will argue that for such a small region, there is insufficient entanglement for a description in terms of a smooth emergent spacetime to be valid.
Correspondingly, we will find that the time evolution in our quantum circuit is trivial before the time $t_1$, the ``time of one e-bit'', when our comoving patch contained just one pair of entangled degrees of freedom, and had size $\approx \ell_p$.
Before this time,
the space corresponding to our comoving patch had not yet emerged.

We will argue that the total number of $e$-folds of inflationary and post-inflationary expansion since the time $t_1$ is bounded,
\be \label{thebound}
N_\textrm{tot} \leq 140\,.
\ee
This implies an upper bound on the total number of $e$-folds of inflation between the time $t_1$ and the time of reheating,
which is close to (but safely above) the number needed to provide a resolution to the horizon problem \cite{Guth:1980zm,Starobinsky:1980te,Brout:1977ix,Sato:1980yn}.
(The upper bound (\ref{thebound}) is related to other bounds that rely on the finite dimensionality of Hilbert space \cite{Banks:2003pt,Kaloper:2004gp,Lowe:2004zs,ArkaniHamed:2007ky,Phillips:2014yma},
as we explain
in the Discussion.)

One aim of this note is to initiate a new approach to the trans-Planckian problem. The time $t_1$ entering the bound (\ref{thebound}) is the time when our comoving region was one Planck length across, and correspondingly when modes that are horizon-sized today had Planckian wavelengths.
Although our quantum circuit has trivial time evolution before this point, it provides a simple quantum-mechanical model for the time \emph{shortly after} $t_1$, when a geometric description of our comoving region was not yet valid, but ancilla qubits were beginning to become entangled, as a precursor to the emergence of a smooth geometry.  One might therefore adapt our setup to examine the quantum state of the curvature perturbations for $t \approx t_1$.
More broadly, our approach provides a speculative but relatively concrete framework for answering certain questions about the very early history of our universe: what happened when our region was Planckian in size?   Does cosmological expansion proceed continuously or in quantized steps?

\section{The Trans-Planckian Problem}

We will first briefly recall some well-known aspects of the trans-Planckian problem in inflationary cosmology (see \cite{Brandenberger-Martin:2013} for a review).

We consider a flat Friedmann-Robertson-Walker universe with metric
\begin{equation}
ds^2 = -dt^2 +a(t)^2 d\vec{x}^2\,.
\label{rwmetric}
\end{equation}
A comoving volume ${\mathcal C}$ is one that has fixed coordinates $\vec{x}$  over time, while a Hubble volume at a given time $t$ is a ball of physical radius $H^{-1}(t)$, where $H = \dot{a}/a$.
We write $R(t)$ for the physical radius of ${\cal C}$.
The number of $e$-folds between two times $t_1$ and $t_2>t_1$, denoted $N(t_1,t_2)$, is defined as
\begin{equation}
 e^{N(t_1,t_2)}=\frac{R(t_2)}{R(t_1)}.
\end{equation}
The region of primary interest is the comoving volume that coincides now with our Hubble volume, i.e.~the region such that $R(t_0)=H^{-1}(t_0)$, with $t_0$ the present time.  We refer to this region as ``our comoving patch'' or ``our comoving volume'', and denote it by  ${\mathcal C}^H(t)$.

The trans-Planckian problem is a
potential ambiguity that arises if the total amount of expansion that our comoving patch experienced exceeds the minimum number of $e$-folds needed to solve the horizon problem.
The underlying issue is that semiclassical methods may not be valid for describing modes with wavelengths shorter than the Planck length $\ell_p$.  In cosmologies with considerably more than 60 $e$-folds of inflation, some of the modes visible in the Cosmic Microwave Background (CMB) and Large Scale Structure had such sub-Planckian wavelengths at the beginning of inflation \footnote{This trans-Planckian problem, which concerns perturbations with wavelength $\ll \ell_p$, should not be confused with the problem of controlling super-Planckian field displacements $\Delta\phi \gg M_p$.  The former can occur in inflationary models with $\Delta\phi \ll M_p$.}.
To see this, recall that in an inflationary scenario involving just enough $e$-folds to solve the horizon problem, the large-angle modes of the CMB exited the horizon at the very beginning of inflation.  Because these modes now have wavelengths of order $H_0^{-1}\approx e^{140}\ell_p$, with $H_0$ the present-day Hubble parameter, they were sub-Planckian at the beginning of inflation if the total number $N_\textrm{tot}$ of $e$-folds of inflationary and post-inflationary expansion exceeds 140.
The critical number $N_\textrm{tot} = 140$  does not depend on the equation of state during or after inflation, nor on the reheat temperature $T_{RH}$.  However, the division of $N_\textrm{tot}$ into $N_I$ $e$-folds of inflationary expansion and $N_{P}$ $e$-folds of post-inflationary expansion does depend on $T_{RH}$.  (For example, with Standard Model particle content and $T_{RH}=10^{15}\,{\rm{GeV}}$, the critical value of $N_I$ is $75$.)

Thus, in cosmologies with
\begin{equation}
N_\textrm{tot} = 140 + \Delta N
\end{equation}
total $e$-folds of expansion between the beginning of inflation and the present, all modes of present wavelength $\le H_0^{-1}e^{\Delta N}$
were sub-Planckian at the beginning of inflation.  While it may be that such a cosmology can be described semiclassically, with modes in the Bunch-Davies vacuum or in another well-behaved vacuum state, the trans-Planckian question consists of taking seriously the issue of modes with sub-Planckian wavelengths.

Requiring that modes of current wavelength $H_0^{-1}$ had wavelengths longer than $\ell_p$ at the start of inflation is equivalent to requiring that our comoving region, of current size $H_0^{-1}$, had size greater than $\ell_p$ at the start of inflation.
However, in most reasoning about trans-Planckian issues, the physical origin of the ambiguity is the problem of specifying the quantum state of
modes with sub-Planckian wavelengths, not the problem of describing a comoving region of sub-Planckian size.  After all, in classical gravity a comoving region is just some chosen subset of a spacelike hypersurface, and at the level of a classical, homogeneous FRW cosmology there is no obvious problem when the size of this subset becomes sub-Planckian.  Correspondingly, the trans-Planckian problem is usually understood as a question about the initial state of currently-observable modes, rather than as a hard upper bound on the total amount of expansion in our history.
The condition
$N_\textrm{tot} \le 140$ is then a bound only to the extent that one insists on making predictions for the CMB without providing a description of modes of sub-Planckian wavelengths.

Here we will derive a superficially identical upper limit, $N_\textrm{tot} \le 140$, from rather different assumptions.  Importantly, in our treatment $N_\textrm{tot} = 140$ turns out to be a
limit beyond which our description of the \emph{background}, not just of the inflationary perturbations, fails.
Specifically, we will find that for $N_\textrm{tot} > 140$, a description of our comoving region as emerging from entanglement is inapplicable at sufficiently early times.
Thus, inflationary cosmologies with $N_\textrm{tot} > 140$ necessarily violate one or more of our assumptions, which we enumerate below.  Although we believe these assumptions are all plausible, any of them may reasonably be questioned.  Our main point is to explore the consequences of assuming their validity.

\section{Counting Entangled Degrees of Freedom}

Our analysis rests on counting entangled degrees of freedom in de Sitter space, so we will first recall the relevant definitions in a quantum-mechanical toy model without gravity.  Consider $N$ spins in some subsystem $A$, and $N'$ spins in $A^{C}$, the complement of $A$.
(By specifying ``spins'' we imagine that the factorization of Hilbert space into fundamental qubits is fixed, so that the amount of entanglement between any particular degrees of freedom is uniquely defined.)
The entanglement entropy of $A$ across the bipartition is defined as $S_A = -\Tr \rho_A \ln \rho_A$,
where $\rho_A$ is the reduced density matrix of $A$.
The \emph{number of entangled degrees of freedom} in $A$, $n_e(A)$, is defined as the number of qubits in the $A$ subsystem that are nontrivially entangled with at least one other qubit,
either inside or outside $A$.
For bipartite entanglement, the number of entangled degrees of freedom is (up to factor of 2) a quantity known as the distillable entanglement.  There are multipartite generalizations of this quantity as well, though computing them becomes more challenging.

The amount of entanglement in $A$ can also be expressed in terms of e-bits.
An e-bit is a unit of entanglement that corresponds to
the entanglement entropy of one half of a Bell pair.
The number of e-bits is equal to the total amount of entanglement (in the form of Bell pairs or other fundamental units of multipartite entanglement, such as GHZ or W states for tripartite information) that can be extracted from a quantum state through a theoretically optimal distillation protocol.
Said colloquially, the number of e-bits is the amount of entanglement inherent in a quantum state.
Because each degree of freedom can share at most one e-bit of entanglement with the entire remainder of the system, the number of e-bits in the spins in $A$ is bounded above by $n_e(A)$.

For a fixed tensor product decomposition of the Hilbert space into a set of qubits, the number of entangled degrees of freedom is a well-defined property of a quantum state, whereas the entanglement entropy of a subregion depends on how that subregion is defined \footnote{The number of entangled degrees of freedom is not necessarily invariant under a global change of basis implemented by some unitary transformation that can change the notion of a ``fundamental qubit.''  For example, in the AdS/CFT correspondence the entangled boundary degrees of freedom are mapped nonlocally to the bulk. In more specific examples, such as those  illustrated by Exact Holographic Mapping \cite{Qi:2013caa} or MERA as a unitary quantum circuit \cite{Evenbly:2007hxg}, a change of basis turns a highly entangled set of qudits into a set that has little or no entanglement. However, for our purposes, $n_e$ should be understood as being defined with respect to a fixed set of qubits.}.
For example, simply by considering a bipartition where all of the entangled degrees of freedom reside on one side of a bipartition, one makes the entanglement across that bipartition zero, while this choice has no effect on the number of entangled degrees of freedom in the state as a whole.

Suppose the Hilbert space of the spins in $A+A^{C}$ has dimension $D=D_A \times D_{A^C} = 2^{N} \times 2^{N'}$, with $D_A$ and $D_{A^C}$ the dimensions of the subspaces describing $A$ and $A^C$, respectively. Assume, without loss of generality, $D_A\leq D_{A^C}$.
We say that $A$ is maximally entangled with $A^C$ if every spin in $A$ is maximally entangled with one or more spins in $A^C$.
In such a case the entanglement entropy of $A$ across the bipartition is maximized, so that
\be \label{snln2}
S_A=\ln D_A = n_e(A) \ln 2 = N \ln 2\,.
\ee
That is, in maximally entangled configurations, the number of entangled degrees of freedom determines the dimension of the entangled subsector of the Hilbert space.
(In the case where the joint $A$, $A^C$ system is a thermal system, the entanglement entropy is nearly maximized and (\ref{snln2}) is approximately satisfied.)  More generally, however, $n_e$ can be much greater than the entropy. Take, for example, the $N$-party GHZ state, where $n_e=N$ but $S=1$ for any non-trivial bipartition.

\section{Framework and Assumptions}

We now consider the number of entangled degrees of freedom in the context of cosmology.  Our assumptions are as follows:
\begin{enumerate}[i.]
\item The evolution of an approximately homogeneous comoving region of space can be described as that of a density matrix associated with a factor of Hilbert space of fixed, finite dimension $D$.
\item At any time $t$, to a given comoving region ${\mathcal C}$ there is associated a number $n_e(t)$ of entangled degrees of freedom describing the spacetime (and matter) structure of $\mathcal{C}$.  There are also $n_u(t) = n-n_e(t)$ unentangled degrees of freedom, with $n={\rm{log}}_2 D$, which
     play no role in the emergent semiclassical geometry or matter configuration.
\item For a Hubble volume in de~Sitter space with cosmological constant $\Lambda$ and Hubble constant $H_{\Lambda}=\sqrt{\Lambda/3}$, the total number of entangled degrees of freedom is approximately the de Sitter entropy,
\be
  n_e(\dS) \approx \frac{S_\dS}{\ln 2},
  \label{ne-ds}
\ee
We assume that (\ref{ne-ds}) holds to good approximation as long as the expansion is very close to de~Sitter, as it is in inflation and in the present epoch.
\item The number of entangled degrees of freedom $n_e(t)$ in the comoving volume ${\mathcal C}^H$ that now coincides with our Hubble volume can never be less than one.
\end{enumerate}
Let us briefly discuss the motivation for these assumptions.

The first assumption, that the quantum state of our comoving region of space is described by a fixed factor of Hilbert space, is a well-justified approximation in a universe that is nearly homogeneous on large scales.
In a general curved spacetime, it would be hard to think of a given region of space as describing a fixed quantum system for all times, as there is no preferred way to evolve it into the future.
(Equivalently, there is no preferred timelike vector field along which to associate a spatial region at one time with one at other times.)
But the large-scale homogeneity of our observed universe allows us to define comoving coordinates, and to use these to
divide the universe into well-defined regions. (There is now a preferred vector field, orthogonal to the hypersurfaces of homogeneity.)
This is not to say that our comoving patch evolves as a causally closed system; individual photons, for example, certainly enter and leave such a region.
But a photon entering our comoving patch does not represent a new degree of freedom; it is described by previously unexcited degrees of freedom (the vacuum of the electromagnetic field) now becoming excited.
Information is entering our region, in other words, but not new qubits, much like a wave traveling through the ocean is not made of a fixed set of water molecules.
In our picture, the entangling of qubits causes space to expand, but
the qubits themselves were always part of the Hilbert space factor describing our observable patch, and were simply unentangled initially.
Unentangled ``ancilla'' qubits of this sort are commonplace in quantum circuits and tensor networks, including in the description of emergent holographic spaces \cite{Kitaev:2002:CQC:863284,Vidal:2008,Swingle:2009bg,Orus:2013kga,Bao-et-al:2015,Czech-et-al:2015,Pastawski:2015qua,Hayden:2016cfa}.

A crucial feature in our analysis is that the dimension $D$ of the Hilbert space of our comoving region is finite. There are well-known obstacles to imagining that regions of spacetime are described by finite-dimensional Hilbert spaces, including the fact that the Lorentz group does not admit nontrivial finite-dimensional unitary representations.  On the other hand, there is suggestive evidence
--- for example, the Bekenstein/holographic bounds \cite{Bekenstein:1973,Bousso:2014sda,Bianchi-Myers:2012} --- that in quantum gravity the Hilbert space of a finite region is indeed finite, and for our purposes we will accept this as a working assumption.  Note also that reasoning about complementarity leads to similar conclusions \cite{Dyson:2002pf}, although our work does not rely on the validity of complementarity.

The second assumption is inspired by the program of relating spacetime geometry to quantum entanglement (for a review, see \cite{VanRaamsdonk:2016}). While geometry from entanglement was originally motivated from examples in AdS/CFT, more general constructions of emergent spacetime
are possible \cite{Maldacena-Susskind:2013,Cao-Carroll-Michalakis:2016},
including tensor network descriptions in which the entanglement of finitely many quantum degrees of freedom creates connectivity that reflects the geometry \cite{Swingle:2009bg,Evenbly-Vidal:2011}.
It has also been argued that MERA \cite{Vidal:2008} can be interpreted as de Sitter space \cite{Beny:2011,Czech-et-al:2015,Kunkolienkar-et-al:2016}. Entanglement is crucial for such models of emergent space, where smoothness and connectedness of space usually corresponds to a large number of degrees of freedom being entangled in an organized manner. In the case where there is little or no entanglement among the quantum degrees of freedom associated with spatial regions, the spatial geometry becomes disconnected, and in certain contexts a firewall can form.

The third assumption captures the idea that a horizon-sized patch of de~Sitter space is an equilibrium system, with a maximum entropy. The Gibbons-Hawking entropy of a Hubble volume is proportional to the area \cite{Gibbons-Hawking:1977},
\be
S_\dS= \frac{A_\dS}{4\ell_p^2}=\frac{\pi}{\ell_p^2H_{\Lambda}^2}\,,
  \label{ds-eq}
\ee
and can be interpreted as the entanglement entropy across the horizon.
The approximate equality (\ref{ne-ds}) is then a conjectured property motivated by the near-equilibrium character of an approximately de Sitter phase.

The final assumption, that $n_e(t) \geq 1$ for our comoving patch, is crucial to our argument.  This assumption may be less familiar and less plausible than the others, so let us explain the justification.  The statement is that for $t$ such that $n_e(t)<1$, the degrees of freedom in our comoving patch were not entangled with anything.
Because our region of space literally \emph{is} that collection of entangled qubits, there is a sense in which our space did not exist before $t_1$.  Said more carefully, our semiclassical region of space had not yet emerged at such early times: the factor of Hilbert space that would eventually describe our comoving region did not contain even a single e-bit, and there was no geometric interpretation of the degrees of freedom that we find around us today.
Nonzero entanglement is necessary for smooth spacetime.

The quantum circuit of Fig.~\ref{fig:circuit} provides a useful perspective on assumption (iv).
In this setting, time evolution is described via the discretization of the Hamiltonian evolution into smaller unitaries in the form of quantum gates.
When no gates are being applied to the quantum degrees of freedom in the comoving patch, these degrees of freedom are evolving trivially.
However, it is always conceivable that space is infinite in extent, and that the total dimension of Hilbert space (once we include regions outside our observable patch) is infinite.  Our comoving patch is then a subset of a larger circuit that allows other patches or observers to evolve further back in time, to times before $t_1$.
As far as the degrees of freedom in our comoving patch are concerned, the larger ``super-circuit,'' whose degrees of freedom are detached from our own, has no effect whatsoever on our comoving patch, as long as no nontrivial gate acts on the degrees of freedom in our comoving patch. This is indeed consistent with a crude model that uses MERA to model the initial inflationary phase.  After the last entangled degree of freedom of our comoving patch is disentangled from the rest of the network, evolving time further backwards on the remaining circuit will not apply any more gates on the qubits describing our region.
If instead our comoving patch is not a part of a larger circuit, the initial state is necessarily pure and there is no further time evolution backwards to times $t<t_1$.

\section{Upper Bound on Total Expansion}

We will now derive a bound on the amount of inflation that can be described in our quantum circuit model.
Denote by $n_e({\mathcal C}^H,t_I)$ the number of entangled degrees of freedom at time $t_I$ in our comoving volume ${\mathcal C}^H$.
We are interested in finding the critical time $t_1$
(``the time of one e-bit'') when this number barely exceeds 1,
\be
n_e({\mathcal C}^H,t_1)\approx 1\,.
\ee
We denote the physical size $R({\mathcal C}^H)$ at $t=t_1$ by $R(t_1)$.

Reasoning based on holography and black hole physics strongly suggests that $R(t_1) \gtrsim \ell_p$, corresponding to at most of order one entangled degree of freedom per Planckian-sized region.
To see this, consider a black hole with radius of order the Planck scale. It is widely believed that a black hole provides the densest packing of information into a region of a given size, and is also maximally entangled with the remainder of the universe it resides in. A system that is maximally entangled with its purifying region has $S=n_e \ln 2$, as previously discussed. Moreover, the entropy of the black hole is given by the Bekenstein-Hawking formula \cite{Bekenstein:1973}, $S={A}/(4\ell_p^2)$, which the Gibbons-Hawking formula closely parallels.  This suggests that the number of entangled degrees of freedom inside a Planck-sized black hole is also of order unity.
Taking a region of space to have a number of entangled degrees of freedom less than or equal to that of a black hole of comparable size, we conclude that
the number of entangled degrees of freedom in a Planckian-sized region ${\mathcal C}$ obeys
\begin{equation} \label{neineq}
n_e({\mathcal C})\lesssim 1\,.
\end{equation}
Indeed, it has been suggested that the black hole bound, at least for entanglement entropy, is saturated in de Sitter space \cite{Bianchi-Myers:2012}.

We will take (\ref{neineq}) to be approximately saturated, corresponding to $R(t_1) \approx \ell_p$, which leads to the loosest bound on the total duration of inflation: the quantum circuit picture then becomes valid once $R({\mathcal C}^H) \gtrsim \ell_p$.
(We have not excluded the possibility that $R(t_1) \gg \ell_p$, in which case our description becomes valid only when $R({\mathcal C}^H) \gg \ell_p$, but this
would lead to a tighter bound on the duration of inflation.)
We thus find that
\be\label{eq:ourbound}
N(t_1,t_0) \lesssim -\ln(H_0 \ell_p) \approx 140.
\ee
Our argument for this result has been fairly general; in the Appendix we discuss more specific assumptions about de~Sitter entanglement that lead to tighter bounds.

Note that (\ref{eq:ourbound}) bounds the total amount of inflationary and non-inflationary expansion between the time $t_1$, defined by the property that our comoving volume contains a single e-bit at $t=t_1$, and the present time $t_0$, with $n_e(t_0) \sim 10^{122}$.
It does not directly limit expansion occurring before $t_1$ or after $t_0$, but as we discuss further below, our logic does lead to suggestive statements about these times.

Our bound on the total number of $e$-folds of expansion can be converted into one on the number of $e$-folds of inflation using standard cosmology.
If reheating to a temperature $T_{\rm RH}$ is approximated as instantaneous, and occurs at a time $t_{\rm RH}$, the number of $e$-folds since reheating is
\be
  N_{\rm{post}} = \ln\frac{a(t_0)}{a(t_{\rm RH})}\,.
\ee
The scale factor is related to the temperature and the effective number of relativistic degrees of freedom $g_{*S}$ via $a \propto g_{*S}^{-1/3}T^{-1}$.  In the Standard Model, $g_{*S}$ is of order $4$ today, and of order $100$ above the electroweak scale. We therefore have
\begin{align}
  N_{\rm{post}}   &\approx \ln\Big(3 \frac{T_{\rm RH}}{T_{0}}\Big)\\
  & = 65 + \ln(T_{\rm RH}/10^{15}~{\rm GeV})\,.
\end{align}

This expression only depends on the current CMB temperature, $T_0 = 2.25\times 10^{-4}~$eV, and the expected entropy production in the Standard Model; it is independent of the equation of state of the universe since inflation.
If inflation ends near the GUT scale, $T_{\rm RH}\sim 10^{15}~{\rm GeV}$, our limit (\ref{eq:ourbound}) implies that the total number of effective inflationary $e$-folds is bounded by
\be
  N_I = N_{\rm{tot}} - N_{\rm{post}} \lesssim 75\,.
\ee
Lower reheating temperatures lead to weaker bounds on $N_I$, e.g.~for $T_{\rm RH}\sim 10^{5}~{\rm GeV}$ we have $N_I \lesssim 98$.

In a model with a given reheating temperature $T_{\rm RH}$, solving the horizon problem requires
\be
N_I \ge N_{\rm{post}} - \frac{1}{2}\ln(1+z_{eq}) \approx N_{\rm{post}} - 4\,,
\ee
with $z_{eq}$ the redshift of matter-radiation equality.  
Thus, according to our bound (\ref{eq:ourbound}), the number of inflationary $e$-folds since the time $t_1$ can exceed the number needed to solve the horizon problem by at most
\be
N_{\rm{extra}} = 14-2\,\ln(T_{\rm RH}/10^{15}~{\rm GeV})\,.
\ee
The bound (\ref{eq:ourbound}) is therefore compatible with solving the horizon problem via inflation occurring after the time $t_1$, but does not allow for much ``unnecessary'' inflation after $t_1$.

\section{Details of the Quantum Circuit}

A few features of our proposed quantum circuit deserve additional explanation.
First of all, in our setup the initial state contains no entanglement.
This is a highly non-generic situation, but is perfectly compatible with the empirical fact that our universe began in a state of very low entropy \cite{Carroll:2014uoa} -- the von~Neumann entropy of any subregion described by the initial state of the circuit will vanish.
We do not attempt here to provide an explanation for this well-known cosmological fine-tuning, merely to model it.

Second, the circuit in Fig.~\ref{fig:circuit} is constructed to describe our comoving region $C^{H}(t)$, but causal influences have entered $C^{H}(t)$ at various times in cosmic history.
Such influences should be represented by the action of gates that entangle the qubits shown in Fig.~\ref{fig:circuit} with qubits describing different degrees of freedom elsewhere: for example, an atom outside $C^{H}$ might emit a photon that we detect.
A more rigorously complete quantum circuit representing cosmological evolution would include gates describing such processes, which we are neglecting here.
As discussed above, the entry of a a photon into our region does not change the dimensionality of our Hilbert space, though it does change the quantum state of our region.

What matters for our analysis is that the spacetime structure of our comoving region comes into being (interpreted semiclassically as ``the universe expands'') by entangling existing degrees of freedom within our Hilbert space, rather than by attaching additional degrees of freedom from outside.
This is our answer to the questions posed in the introduction about the appearance of new modes as the universe expands.
For cosmological evolution described by a finite-dimensional Hilbert space, the total number of degrees of freedom is always fixed. In essence, the quantum circuit picture presents a natural framework for the newly ``created'' modes to become entangled with the rest when they are no longer trans-Planckian, by modeling the process as ancillary qubits becoming entangled as time evolves.

The circuit picture provides a concrete, operational sense in which the condition that there should be at least one entangled degree of freedom can be made precise. Namely, no information about the spacetime is imparted by gates acting on qubits in our comoving region at times $t<t_1$.
All entanglement that sources such information must be injected by unitaries that appear later in the circuit, i.e.~at a time after the ``beginning'' of the universe.

We can ask how the quantum circuit picture applies to the future evolution of the universe.
One the one hand, it is conceivable that we live in a universe near a late time de Sitter
vacuum in which almost all of the $n=S/\ln{2}$ degrees of freedom in the entire Hilbert space of our region are entangled, and all further time evolution simply increases the circuit complexity.
It may then be possible to adopt the complexity picture where time evolution is directly defined by the growth of complexity \cite{Brown:2015lvg}.

On the other hand, it is also plausible that the current de Sitter phase is metastable, and can decay into a vacuum with a smaller (positive) cosmological constant, and so a larger entropy $S'>S$.  We will not give a quantum circuit description of the associated tunneling process in this work.
However, we remark that $(S'-S)/\ln{2}$ additional ancilla qubits are required, and the gates that entangle them with the degrees of freedom of the false vacuum are different from those that describe exponential expansion within the false vacuum.

\section{Discussion}

We have proposed a quantum circuit picture for cosmological expansion.
Our fundamental assumption was that expansion corresponds to the progressive entanglement of degrees of freedom that were initially unentangled.
Time evolution corresponds to the application of quantum gates that create entanglement,
and the amount of cosmic time elapsed since the unentangled initial state is determined by the circuit complexity.
We posited that the total number of degrees of freedom in our Hubble volume is a finite number related to the area of the de~Sitter horizon, and that the number of entangled degrees of freedom in a region with a geometric description can never be less than one.

This picture differs markedly from the conventional intuition based on quantum field theory in curved spacetime. In that context, the number of degrees of freedom (and hence the dimensionality of Hilbert space) is potentially infinite. In an effective description with a Planck-scale cutoff, new degrees of freedom are continually
appearing as they expand from sub-Planckian to safely super-Planckian wavelengths. Our picture seems more compatible with the principles of unitary evolution (new degrees of freedom are never created) and holography (the total number of degrees of freedom is finite in a de~Sitter universe).

In our approach, there is a general upper bound on the number of $e$-folds of cosmological expansion since the time $t_1$ when there was a single pair of entangled degrees of freedom in the comoving region that now coincides with our Hubble volume.
The number of inflationary $e$-folds consistent with this bound is comfortably, but not parametrically, larger than what is needed to solve the horizon problem.

Our bound limits the total number of $e$-folds of expansion within our comoving patch, but our ignorance about the underlying theory of quantum gravity allows for potentially different global scenarios.
If we take seriously the de~Sitter entropy as telling us the dimensionality of the Hilbert space corresponding to our observable patch of spacetime, there are two possibilities. One is that this Hilbert space represents the entire universe; there is no larger multiverse described by additional degrees of freedom, and only the degrees of freedom in the bulk and on the boundary of de~Sitter exist \cite{Banks:2000fe,Witten:2001kn,Parikh:2004wh,Nomura:2011rb}. In that case our bound is a straightforward limit on the total amount of expansion space can undergo before reaching its de~Sitter equilibrium state.

The other possibility is that our observable patch represents only part of the universe, and its Hilbert space is just one part of a larger Hilbert space. In that case,
the classical universe is much larger than what we observe. It follows that there could be many more $e$-folds of total expansion than what our bound indicates. However, even in that case our bound applies to the number of physically meaningful $e$-folds of expansion of our own space.
Any additional expansion occurring before $t_1$ did not involve any of the degrees of freedom that currently constitute the spacetime geometry in our observable universe: at that early stage our degrees of freedom were completely unentangled, and the space that was then expanding now corresponds to regions outside our comoving volume.
In this sense, our bound applies to the universe we see, even if the full theory describes additional degrees of freedom as well.

Our upper bound on inflationary $e$-folds is similar to a bound derived by Banks and Fischler  \cite{Banks:2003pt}; see also \cite{Kaloper:2004gp,Lowe:2004zs,Phillips:2014yma}.
Like ours, their bound comes from assuming that physics in a de~Sitter patch is described by a finite-dimensional Hilbert space.
Unlike us, they require that the quantum state be pure rather than mixed and that the de~Sitter phase be absolutely stable, and their early-time constraint comes from insisting that physics be described by an effective quantum field theory, rather than insisting that the entangled spacetime structure contain at least one qubit.
Our bounds are also numerically different, both because we do not invoke any assumptions about the equation of state at early times, and because we find a different scaling of the entropy with the number of $e$-folds.
The bound due to Kaloper, Kleban, and Sorbo \cite{Kaloper:2004gp} is based on what an observer in de~Sitter can conceivably measure, rather than on unitary quantum evolution by itself.
That of Phillips, Scacco, and Albrecht \cite{Phillips:2014yma} is derived within the framework of a de~Sitter equilibrium picture \cite{Albrecht:2011yg}, which again involves a slightly different set of fundamental assumptions.
The spirit behind these various bounds is certainly similar; we believe that the one presented here is based on a simple set of explicit assumptions, and is unique in making direct reference to ancilla degrees of freedom gradually becoming entangled as space expands, but our logic is not incompatible with that of previous bounds.

There is also related work by Arkani-Hamed \emph{et al.}, which proceeds from weaker assumptions than those we have invoked, and finds bounds exponentially weaker than our own \cite{ArkaniHamed:2007ky}.  Their analysis is different: they rely on the slow increase of entropy during slow-roll inflation, which follows from the gradual reduction of the energy density during that phase.
They then obtain a bound involving the de Sitter entropy at the end of non-eternal inflation, which we may denote $S_{\rm{dS}}^{\rm{early}}$.
Their logic is to place an upper bound on the number of modes
detectable by a hypothetical future observer in a late-time phase with negligible cosmological constant, e.g. Minkowski space.  We have instead considered an observer in a cosmology that has entered, or is entering, a late-time de Sitter phase (possibly but not necessarily metastable)
with finite cosmological constant $\Lambda>0$, and finite de Sitter entropy $S_{\rm{dS}}^{\rm{late}}$.  Notice that our analysis allows $N_{\rm{tot}} \to \infty$ in the limit $\Lambda\to 0$.
The bound of \cite{ArkaniHamed:2007ky} (see also \cite{Dubovsky:2011uy}) is
\be
N_{\rm{tot}} \le \frac{1}{12}\, S_{\rm{dS}}^{\rm{early}}\,,
\ee
while ours is
\be
N_{\rm{tot}} \le \ln\left(S_{\rm{dS}}^{\rm{late}}/\pi\right)\,.
\ee
Thus, our bound is compatible with, but quite different from, that of \cite{ArkaniHamed:2007ky}.

In closing, let us point out some possible applications of our picture.
One advantage of describing time evolution through a quantum circuit is that one can in principle reverse the computation. As the quantum circuit we outlined in this work is a unitary circuit, one can imagine simply running it in reverse, with generic data about the quantum state today, to gain intuition about what generic states in the early universe could have looked like. This could be done either via Monte Carlo generation of the state of the current universe, or through some ansatz for the current entanglement. (Our circuit is meant to describe the full quantum state of the universe, not only the branch of the wave function we find ourselves on; details of the actual state are therefore unobservable to us.)
This computation would of course require knowledge of the specific gates in the circuit.
Performing the evolution backward in time by inverting each gate separately is in principle much simpler than inverting a generic unitary operator acting on the Hilbert space.

We stress that although the spacetime picture breaks down when the number of entangled qubits is not large, the circuit as a unitary picture does not.  By an appropriate modification of the quantum circuit presented here one could aim to explore the initial state of inflationary perturbations, by characterizing how the degree of entanglement of the state evolves as a function of time.
More ambitiously, entanglement in a phase where $n_e(t)$ is not large (but $n_e(t)>1$) might provide a model for the chaotic conditions at nearly-Planckian densities, and for the emergence of inflating regions in this era.
However, realizing these applications would require the development of a more detailed dictionary between entanglement and (quasi) de Sitter spacetimes.

The assumptions leading to our proposal are not secure beyond reasonable doubt, although they do
seem to follow from plausible conjectures about holography and unitarity.
Perhaps the most important lesson from this analysis is that phenomena in quantum gravity can be very different from our semiclassical intuition, in ways that can have important consequences for cosmology.

\section*{Acknowledgements}
We thank Don Page and Jason Pollack for comments on the draft. N.~B.~would like to acknowledge Sabrina Pasterski and Stephen Jordan for comments on the draft. L.~M.~is indebted to Tom Hartman for extensive explanations, and to Thomas Bachlechner, Sandipan Kundu, and John Stout for helpful discussions and comments on a draft.   L.~M.~thanks the Caltech theory group for their hospitality during the initial stages of this work. C.~C. and S.~C. would like to thank Aidan Chatwin-Davies for detailed comments and discussions on the draft. This research is funded in part by the Walter Burke Institute for Theoretical Physics at Caltech, by DOE grant DE-SC0011632, by the Foundational Questions Institute, by the Gordon and Betty Moore Foundation through Grant 776 to the Caltech Moore Center for Theoretical Cosmology and Physics, and by NSF grant PHY-131622.

\section*{Appendix}

We arrived at the bound $N(t_1,t_0)\leq 140$ by means of relatively general assumptions.
In this appendix we comment on how specific models for the emergence of de Sitter space from entanglement could lead to more restrictive bounds.
These speculations are largely based on \cite{Cao-Carroll-Michalakis:2016}.

Two persistent difficulties in understanding de Sitter space from entanglement are the absence of a spatial boundary and the presence of de Sitter entropy.
This entropy, and hence
the number of entangled degrees of freedom (since de~Sitter is supposed to be an equilibrium state),
scales as the horizon area, whereas local bulk quantities naively scale as volumes. It therefore appears that one cannot assign quantum degrees of freedom locally to each subregion in de Sitter space, as many of these degrees of freedom would have to be shared non-locally to account for the sub-extensive scaling.

On the other hand, in order to be consistent with the picture presented in \cite{Cao-Carroll-Michalakis:2016,Faulkner-et.al.:2013, Jacobson:1995ab,Jacobson:2015}, the entanglement entropy will also need to satisfy an approximate area law for local regions in order for Einstein gravity to emerge. This constrains the kind of structure we are allowed to consider in order to build up de Sitter space. For small regions, the entanglement has to be predominantly short-ranged, and satisfy an area law. However, on longer scales comparable to the Hubble radius, the dominant entanglement will be long-ranged, so that some of the quantum degrees of freedom that encode the geometry of such regions are shared non-locally. This picture of entanglement is similar to that presented in \cite{Verlinde:2016toy}.

To start, we assume that the state from which de Sitter emerges allows us to define subregions of the emergent geometry.
Consider a subregion $A$ of linear size $R$ in a Hubble patch of a de Sitter phase with Hubble constant $H$. We assume that the number $n_e(A)$ of quantum degrees of freedom that encode the geometric information of $A$ is well-defined (even if one cannot necessarily localize all $n_e(A)$ degrees of freedom to the subregion), and that
\begin{equation}
 n_e(A)\approx f_{R}(x) n_e(\dS)\,.
\end{equation}
Here we have some function $0\leq f_R(x)\leq 1$, where $0\leq (x=R H)\leq 1$ parametrizes the dimensionless ratio between the size of the subregion and that of the Hubble patch. The $R$-dependent functional form captures the transition of the scaling behaviour for the number of quantum degrees of freedom, which changes from volume law scaling to area law scaling when considering larger and larger regions.
Correspondingly, the dominant form of entanglement in these regions transitions from short-ranged to long-ranged.

As such, it is reasonable to conjecture that
\begin{equation}
 x^3\leq f_R(x)\leq x^2,
 \label{eqn:assumption}
\end{equation}
which captures the area-to-volume transition.
Now we can derive a bound on the duration of inflation.  Let $x=R(t_1)/R(t_I)$, where $t_1$ marks the beginning of inflation, $t_I$ is any time during the inflationary phase, and $t_0$ is the present.
Then we have
\begin{equation}
n_e(t_1)\approx f_{R(t_1)}(x) \frac{\pi R(t_I)^2}{ \ell_p^2 \ln(2)}\leq x^2\frac{\pi R(t_I)^2}{ \ell_p^2 \ln(2)}=\frac{\pi R(t_0)^2e^{-2N(t_1,t_0)}}{\ell_p^2\ln 2},
\label{eqn:nebound}
\end{equation}
where we have used $R(t_I)\sim H(t_I)^{-1}=H(t_1)^{-1}$.
Combining (\ref{eqn:nebound}) and assumption (iv), we have
\begin{equation}
N(t_1,t_0)\leq \frac 1 2 \log\frac{\pi R(t_0)^2}{\ell_p^2 \ln(2)}\approx 140.
\end{equation}
However, if the form of $f_R(x)$ were known, e.g.~from a model of a de Sitter tensor network, the transition function could be explicitly evaluated.
For instance, suppose we obtained $f_R(x) = x^{q(R)}$ for some $q(R)$ such that (\ref{eqn:assumption}) is satisfied.  Then we could write
\begin{equation}
N(t_1,t_0)\leq \frac{2-q}{q}\ln \frac{R(t_I)}{\ell_p} + \ln \frac{R(t_0)}{\ell_p} +\frac 1 q\ln \frac{\pi}{\ln 2}\approx 140 -\frac{q-2}{q} \ln \frac{R(t_I)}{\ell_p}.
\label{eqn:transition_bound}
\end{equation}
which is sensitive to the inflationary scale and to the entanglement structure of de Sitter. Because $R(t_I)\geq \ell_p$ by (iii) and (iv), we see that for $q(R)$ between $2$ and $3$, (\ref{eqn:transition_bound}) yields a potentially tighter bound.

For instance, following the arguments in \cite{Verlinde:2016toy}, let us assume a $q=3$ relation. Then we find that
\be \label{eq:ourboundcubic}
N(t_1,t_0) \approx -\ln(H_0 \ell_p) + \frac{1}{3} \ln(H(t_1)\ell_p)+\frac{1}{3}\ln \left(\frac{\pi}{\ln 2}\right).
\ee

Observational upper limits on primordial tensor modes \cite{Array:2015xqh} give an upper bound $H(t_{CMB})\ell_p \le 5 \times 10^{-6}$ on the Hubble scale at the time $t_{CMB}$ when the modes visible at large angular scales in the CMB exited the inflationary horizon.  Because $t_1 < t_{CMB}$ in general, we cannot directly bound $H(t_1)$ from observations: the inflationary energy could have diminished noticeably between $t_1$ and $t_{CMB}$.  But if we could exclude a rapid decrease in energy over that interval --- perhaps through limits on the scale-dependence of the scalar and tensor power spectra --- and so have $H(t_1) \sim H(t_{CMB})$, the bound (\ref{eq:ourboundcubic}) would read
\be  \label{eq:ourboundobs}
N(t_1,t_0) \lesssim 140+\frac{1}{3} \ln(5\times 10^{-6}) = 136\,.
\ee

For inflation that ends near the GUT scale, $T_{\rm RH} \sim 10^{15} \rm GeV$, we would have $N_I\lesssim 71$. Lower reheating temperatures lead to weaker bounds on $N_I$, e.g. for $T_{\rm RH} \sim10^5 \rm GeV$ we have $N_I \lesssim 98$.

\bibliography{circuit_inflation}
\bibliographystyle{JHEP}


\end{document}